# Vibration and jitter of free-flowing thin liquid sheets as target for high-repetition-rate laser-ion acceleration

Zhengxuan Cao[1], Ziyang Peng[1], Yinren Shou[1,4], Jiarui Zhao[1], Shiyou Chen[1], Ying Gao[1], Jianbo Liu[1], Pengjie Wang[1,5], Zhusong Mei[1], Zhuo Pan[1], Defeng Kong[1], Guijun Qi[1], Shirui Xu[1], Zhipeng Liu[1], Yulan Liang[1], Shengxuan Xu[1], Tan Song[1], Xun Chen[1], Qingfan Wu[1], Xuan Liu[1], Wenjun Ma[1,2,3*]

[1]State Key Laboratory of Nuclear Physics and Technology, and Key Laboratory of HEDP of

the Ministry of Education, CAPT, Peking University, Beijing 100871, China

[2]Beijing Laser Acceleration Innovation Center, Huairou, Beijing 101400, China

[3]Institute of Guangdong Laser Plasma Technology, Baiyun, Guangzhou 510540, China

[4]Present address: Center for Relativistic Laser Science, Institute for Basic Science, Gwangju 61005, Korea

[5]Present address: Institute of Radiation Physics, Helmholtz-Zentrum Dresden-Rossendorf, Dresden 01328, Germany

* wenjun.ma@pku.edu.cn

**\* Correspondence:**
Wenjun Ma
wenjun.ma@pku.edu.cn

**Keywords: Liquid sheet target, Laser-ion acceleration, High-repetition-rate targets, Hydrodynamic instability, Rayleigh length**

## Abstract

Very thin free-flowing liquid sheets are promising targets for high-repetition-rate laser-ion acceleration. In this work, we report the generation of micrometer-thin free-flowing liquid sheets from the collision of two liquid jets, and study the vibration and jitter in their surface normal direction. The dependence of their motion amplitudes on the generation parameters is studied in detail. The origins of the vibration and jitter are discussed. Our results indicate that when the generation parameters are optimized, the motion amplitudes in the stable region can be stabilized below 3.7 μm to meet the stringent requirement of sheet position stability for a tight-focusing setup in laser-ion acceleration experiments.

## 1   Introduction

The interaction of ultraintense laser pulses with solid targets at relativistic intensity ($10^{18}$ W/cm$^2$) have produced energetic protons close to 100 MeV(1) and heavy ions over 1 GeV(2). Such a novel ion acceleration method based on relativistic laser-plasma interaction attracts extensive attention in the field of particle accelerators. It not only offers an alternative way to compact accelerators but also exceptionally delivers ion beams of unique properties such as small source sizes (micrometer scale)(3, 4), short time durations (picosecond to nanosecond)(5), and ultrahigh peak flux ($10^{10}$

A/cm). Applications of the accelerated ions, including FLASH radiotherapy(6), neutron generation(7, 8), radiolysis chemistry(9), and ultra-fast ion imaging(10), are being extensively studied.

Many potential applications of laser-driven ions require a high-repetition-rate operation (kHz or higher) to provide a high average flux of ion sources. Conventional targetry, where solid thin foils self-support on apertures that are closely arranged on a frame for shooting in sequence, can hardly meet the need(11). Typically, each laser shot would ablate a target, and a new target has to be moved in by motorized stages before the next laser shot. Several challenges arise in these processes. Firstly, solid foil targets are often manually fabricated in advance and cannot be mass-produced in a short time. However, continuous operation at kHz-repetition-rate would consume millions of targets in just one hour. The gap between target preparation and consumption is huge. Secondly, the target frame has a limited surface area. Each frame can carry no more than a thousand targets generally. It can only last for 1 second at 1 kHz repetition rate. Lastly, the less than 1 millisecond interval for target update would also be highly challenging for motorized stages to maintain micrometer-scale accuracy in target positioning.

Alternatively, free-flowing thin liquid sheets(12-23) are promising targets for high-repetition-rate laser-ion acceleration. Firstly, they are self-supporting like solid foil targets, and their electron densities are close to those of solids. Thus, they can be ideal substitutes for solid foils in laser-ion acceleration. Secondly, liquid sheet targets solve the issues of fast fabrication and target update. They are self-renewed under continuous liquid supply. After each shot, the sheet is recovered in less than 0.1 ms, which enables a shooting rate of 10 kHz or higher(15). Thirdly, the flow rates required to generate a liquid sheet are mostly less than 5 mL/min. A typical reservoir is several liters, which can support at least hours of continuous operation without downtime even without recycling. Due to the above advantages, liquid sheets have received more and more attention.

There are two schemes reported to produce very liquid sheets suitable for laser ion acceleration. The first is to collide two liquid jets at a certain angle(12-19), as shown in Figure 1A. A leaf-shaped planar liquid sheet will spontaneously form under laminar flow conditions. The underlying physics is as follows: the momentums of the two colliding jets are redistributed in the orthogonal plane; then the liquid radially expands outward, and gradually retracts under the pulling of surface tension to form a closed liquid sheet. The second method uses a single-exit convergent nozzle(20-23). The cross-section area of the channel gradually decreases till it reaches the minimum value at the exit, creating an equivalent collision effect of liquid inside the nozzle. After ejected from the nozzle exit, the liquid gets rid of the constraint of the channel and expands to form a sheet.

In laser-ion acceleration, the cut-off energies of ions scale up laser intensity I with specific powers depending on the acceleration scheme ($I^{0.25} \sim I^{0.6}$ for target normal sheath acceleration(24)). So the laser pulses are generally tightly focused for higher intensity, where the Rayleigh lengths are less than 10 μm. Due to hydrodynamic instabilities and other factors, the liquid sheets themselves vibrate and jitter in their surface normal direction. If the amplitudes of their motion exceed the Rayleigh length, the results of laser-ion acceleration will be drastically impaired. Therefore, studying the sheets' motion under different generation parameters is necessary for a stable output of ions. Previous studies have reported a general result of $\sigma = 2$ μm by exploiting side-on microscope imaging(14, 15), where $\sigma$ is the standard deviation of sheet positions. However, to our knowledge, studies that extensively investigate the influences of various parameters on surface motion amplitudes based on accurate measurements have not been reported.

In this paper, we perform a systematic study on the vibration and jitter of free-flowing liquid sheets for laser ion acceleration. Their motions were measured by a confocal displacement detector with high precision at the kHz sampling rate, which enabled us to obtain accurate and detailed motion trajectories. We then studied how the generation parameters affect the motion amplitudes of sheets. Summaries of the results and the analysis of the origins of the vibration and jitter are given. In the end, we discussed the feasibility of using liquid sheet targets for laser ion acceleration with a tight focus.

## 2 Methods

### 2.1 Liquid sheets generation

The liquid sheets were generated with our homemade liquid-sheet-generation system. It is composed of a liquid-delivery subsystem (Figure 1B) and a capillary positioning subsystem (Figure 1D). The liquids used in this work were pure ethylene glycol (EG) or aqueous solutions of EG at different mass concentrations. The solutions were prepared by weighing with a balance and confirmed by a portable refractometer. In the liquid-delivery subsystem, the liquids were driven by a high-performance liquid chromatography (HPLC) pump (Shimadzu, LC-20ADXR) with a maximum flow rate of 5 mL/min. It was then equally injected into two symmetrical arms. A homemade pulsation dampener was installed to attenuate the pulsation of flow coming from the pump before the liquid entered the capillaries. The inner diameter of the capillaries was 50 μm. The two liquid jets out of the capillaries collided at a certain angle to produce the liquid sheets. As shown in Figures 1A, 1C, when the jets are in the xOz plane, the resulting liquid sheet lies in the yOz plane. The minimum thickness of liquid sheets was measured as 1.1 μm by white light interferometry (the detail for the measurement is beyond the scope of this work). The lengths of the jets L (defined as the length between a capillary tip and collision point of the two jets, see Figure 1A) and their collision angle ($2\theta$) were controlled by the capillary positioning subsystem (Figure 1D). The lengths of the jets were altered by sliding the mounting bases of the capillaries on the rails. The angle between the jets can be changed by adjusting the angles of the rails.

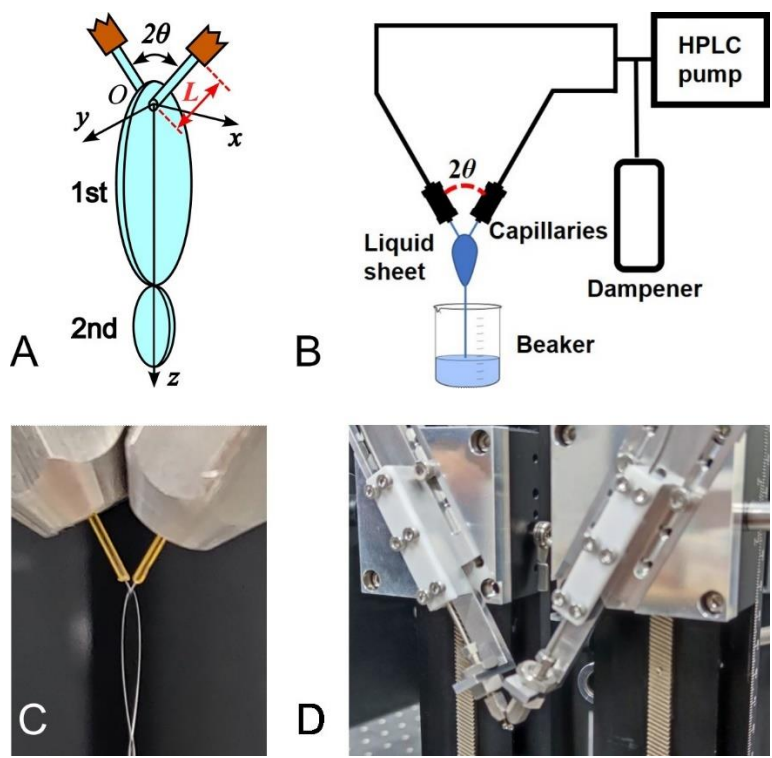

Figure 1. (**A**) Schematic of the liquid sheets generation process in the two-jet-collision regime. (**B**) Schematic of our liquid-delivery subsystem. (**C**) Photograph of the generated liquid sheet. (**D**) Photograph of the capillary positioning subsystem.

## 2.2   Measurements of sheets' motions

We employed a confocal displacement detector (Micro-Epsilon, confocalDT 2421) to measure the motion of the liquid sheets. Figure 2 depicts the setup. The probe head of the detector was placed along the normal direction of the sheet, i.e., the x-direction. The working principle of the confocal displacement detector is as below: white light is focused by a chromatic lens in the sensor so that the focal spot positions for lights with different wavelengths are different; the detector collects the light reflected from the surface of the object (the liquid sheet in our case) and measures its wavelength; according to the measured wavelength, the position of the object can be determined with a 24-nanometer precision. The sampling rate of the detector was set to 1 kHz, which is enough to capture the high-frequency motion of the sheets. In addition to the confocal sensor, there were two lens-CCD imaging systems on the opposite and lateral sides to record the coordinates of measurement points and the lengths of the jets, respectively. Multiple light sources were used for imaging illumination.

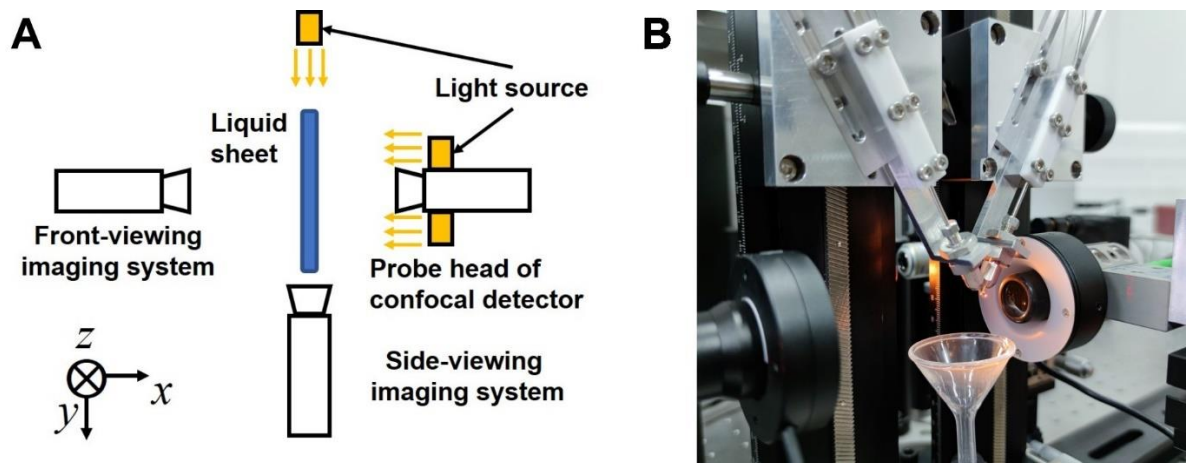

Figure 2. (**A**) Schematic and (**B**) photograph of the measurement setup. The confocal detector is used to measure the motion of the sheets. Two imaging systems record the experimental parameters. Multiple light sources are used for imaging illumination.

## 2.3 Data processing methods

We illustrate the data processing methods and typical measurement results under a set of generation parameters, which is referred to as the control group in this work. The parameters are: pumping flow rate of 3.5 mL/min, the mass concentration of EG 100%, the lengths of the jets 0.9 mm, the pulsation dampener disabled (not in use), and the collision angle ($2\theta$) 60°. There were 7 measurement points along the central line of the sheet (shown in Figure 3A). At each point, the measurements were performed five times. The duration of each measurement was 30 s. Figure 3B shows the raw data of one measurement at a certain point, and Figure 3C is a zoom-in of Figure 3B. One can see that there is high-frequency vibration superposed on low-frequency jitter. The vibration generally accounts for a small part (~20%) of the total amplitude, while the jitter contributes a majority (~80%). For a quantitative study, the inset in Figure 3D presents the histogram of the sheet's position corresponding to one measurement. We define the amplitude of the motion by the threshold $X_{95}$ (below which 95% of values are contained). That is to say, under 95% of circumstances, the position of the sheet will be within $\pm X_{95}$ range. The $X_{95}$ of five measurements for the same measurement point are averaged, yielding $\overline{X_{95}}$ as the motion amplitude at this point.

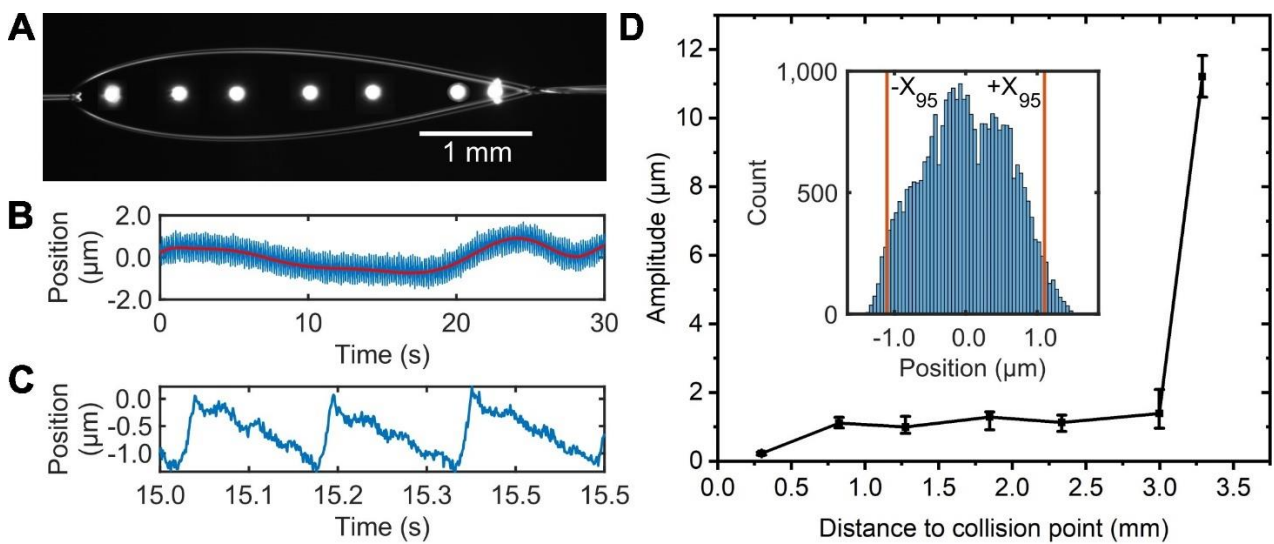

Figure 3. (**A**) The arrangement of the measurement points in a typical experiment. (**B**) Raw data of the motion of a liquid sheet corresponding to one measurement at a certain point. (**C**) A zoom-in of (**B**). (**D**) The motion amplitude of the sheet as a function of the distance to the collision point. The inset in d is the typical histogram of one measurement, where two vertical lines indicate the threshold $X_{95}$.

The motion amplitude of the sheet as a function of the distance to the collision point is shown in Figure 3D. One can see that it's neglectable at 0.25 mm, and is almost constant from 0.8 mm to 3.0 mm with amplitudes around 1 μm. The 2.2-mm-long plateau region is long enough and very beneficial for a stable laser-target interaction. Therefore, this region is defined as stable region of the sheet in this work. We speculate that such small motion amplitudes are due to two reasons. Firstly, the hydrodynamic instabilities (such as Plateau-Rayleigh Instability(25)) are well-suppressed. One can calculate the Reynolds number(15) is only 52, considering the EG's viscosity of 16 mPa · s (under the condition of 25°C and 1 atm) (26), and the small flow rate (3.5 mL/min). The development of the hydrodynamic instabilities for such a small Reynolds number is very slow. Secondly, the vibration of the capillaries is also well-controlled in our case due to the mechanical robustness of the capillary positioning subsystem (Figure 1D). Otherwise, the motion amplitude will increase with the distance to the collision point.

As shown in Figure 3D, the motion amplitude abruptly rises to 11 μm at 3.3 mm away from the collision point. Around this point, the liquid in the first sheet is retracted by the surface tension and starts to form the second orthogonal sheet (as shown in Figures 1A, 1C). In this region, the streamline changes drastically, and it is reasonable to cause greater instability and hence the motion amplitude.

## 3 Results

The motion amplitude of the liquid sheets could be influenced by sheet generation parameters such as the flow rate of pump, the mass concentration of EG solution, the lengths of the jets, the status of pulsation dampener, and the collision angle $2\theta$. We made a systematic study to quantify their influences and present the results below. The data processing methods are the same as those in the control group.

### 3.1 Flow rate

We used four different flow rates (3.5, 4, 4.5, 5 mL/min) while keeping the other parameters the same as those of the control group. It can be seen from Figure 4A that the amplitude monotonically increases with the increased flow rate. This trend is consistent at all three measurement points. We speculate that these phenomena result from the increased reciprocating frequency and driving pressure of the pistons adapted for larger flow rates. These two factors lead to an evident increase in the pump's output power. Hence the mechanical vibration power and flow pulsation transmitted to the liquid sheet are increased.

### 3.2 Mass concentration of EG solution

We performed the experiment with four different mass concentrations (100%, 75%, 50%, 25%) while the other parameters remained the same as those of the control group. Figure 4B illustrates that the amplitude goes down first and then up with the decreased concentration. This trend is the same for all three measurement points. The reason behind this is speculated to be the competition of two factors: the pump's output power and the hydrodynamic instabilities. Decreasing the EG concentration is in fact decreasing the viscosity of the liquid. For a lower viscosity, the pump pressure is lower at a given flow rate and hence the pump's output power. For reference, the pump pressures of the four liquids are ~35 MPa, ~19 MPa, ~10 MPa, and ~5.3 MPa, respectively. Since the liquid is driven at lower pressure and power, the mechanical vibration and flow pulsation transmitted from the pump to the sheet are reduced. When the concentration further decreases to 25%, the viscosity is too low to suppress the growth of the hydrodynamic instabilities. As a result, the motion amplitude of the sheet rises again. Nevertheless, amplitudes at these concentrations are still smaller than the pure EG.

### 3.3 Lengths of the jets

The lengths of the jets took six values (0.9, 3.4, 5.9, 8.4, 10.9, 13.4, 15.9 mm), while the other parameters were the same as that of the control group. Figure 4C shows the amplitude of the sheets as a function of the lengths of the jets. The most notable feature is that the amplitude is very large when the lengths are 10.9 mm. We speculate that it coincidentally corresponds to a resonance mode of the sheet. Besides this abnormal point, the amplitude increases with increased lengths of the jets. This is because the instabilities have a longer time to develop in a longer jet. In fact, we found in the experiments that if the lengths of the jets are longer than 8.4 mm, the developed instabilities will cause the sheet to disintegrate into a spray near the end of the sheet.

### 3.4 Pulsation dampener

The liquid was pumped by the piston with a sinusoidal reciprocating motion, so the instantaneous output flow rate is also pulsing, which is a source of the disturbance. We use a homemade stainless-steel cylinder full of air at 1 atm as the pulsation dampener to attenuate the flow pulsation. Since air is easier to be compressed than liquid, the dampener takes in the excessive liquid at the peak of the flow and sends some liquid back at the valley, and therefore attenuates the flow pulsation. Accordingly, the stimulated instabilities are reduced, and so is the motion amplitude of the sheet. We studied the performance of the dampener while keeping the other parameters the same as that of the control group. As we can see from Figure 4D, after the dampener is enabled, the amplitude decreases by about 0.5 μm in the stable region.

### 3.5 Collision angle (2$\theta$)

Figure 4E shows the motion amplitude of the sheets with collision angles of 60° and 90°, keeping the other parameters the same as that of the control group. It can be seen that when the angle of collision increases from 60° to 90°, the amplitude decreases by about 0.51 μm in the stable region. The previous study has found that increasing the collision angle is detrimental to the stability of the sheets, especially when using low-viscosity liquids at high flow rates(27). However, our results indicate that increasing the collision angle can still be an option to reduce the amplitude if the flow rate is relatively low (3.5 mL/min) and liquid viscosity is high (16 mPa · s).

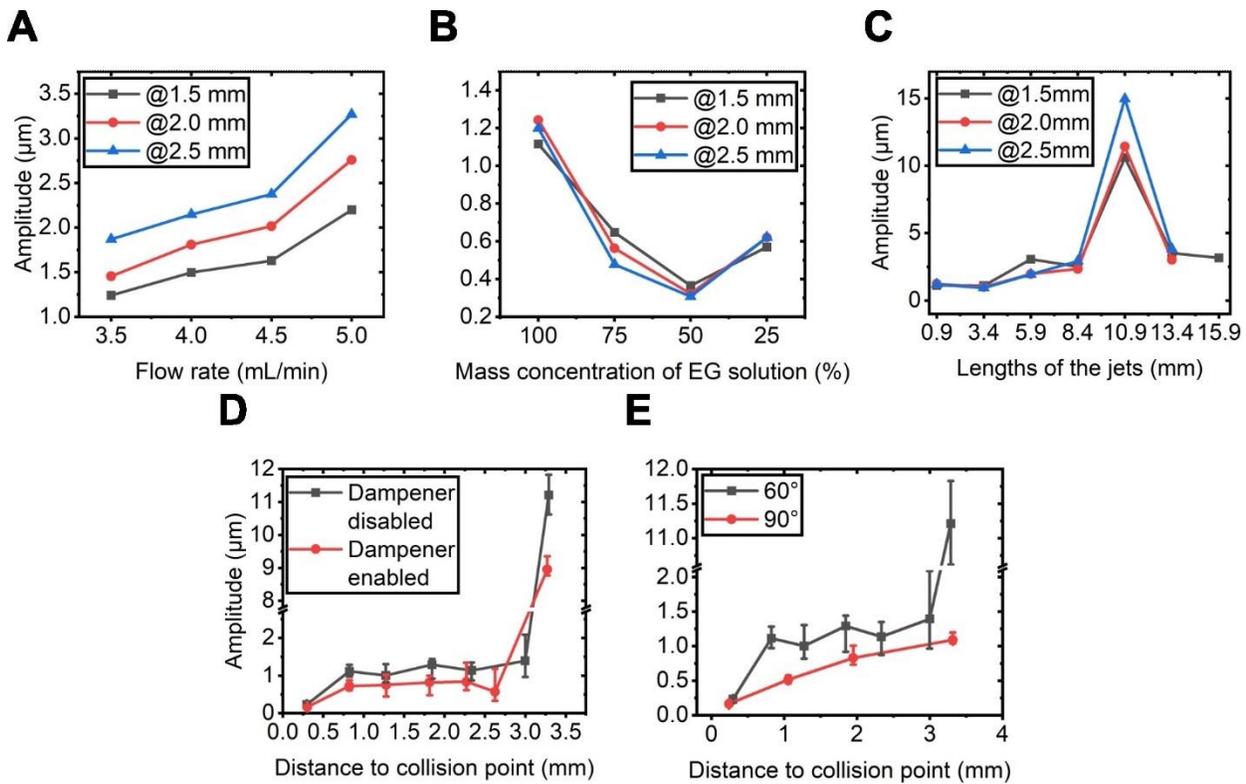

Figure 4. Motion amplitude of liquid sheets with different generation parameters: (**A**) flow rate, (**B**) mass concentration of EG solution, (**C**) lengths of the jets, (**D**) status of pulsation dampener, (**E**) collision angle ($2\theta$).

## 4 Discussion and conclusion

### 4.1 Discussion

For a clear illustration of the influences from various parameters, we summarize all the above results in Figure 5. The data for a specific sheet generation parameter comes from three measurement points with distance to the collision point of 1.5 mm, 2.0 mm, and 2.5 mm, respectively.

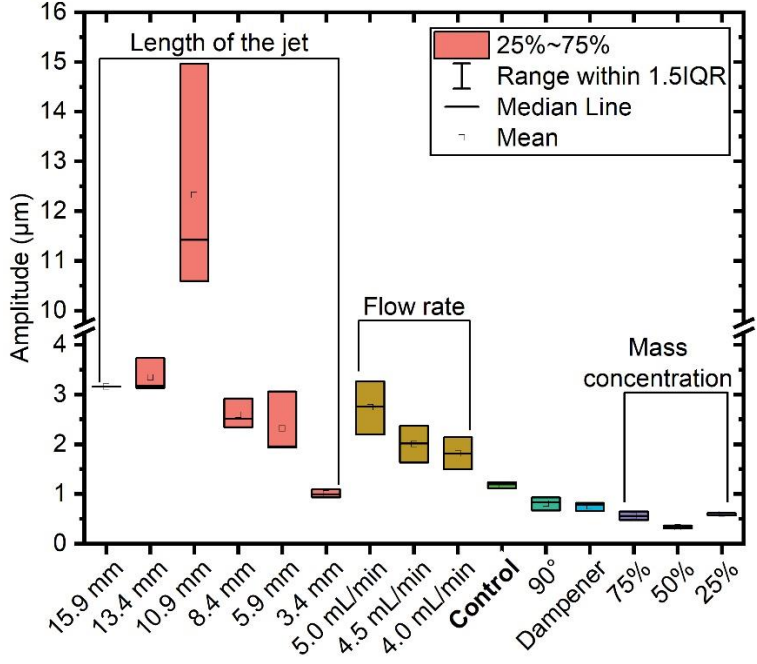

Figure 5. A summary of sheet generation parameters' influences on the motion amplitude of the liquid sheet.

According to Figure 5, in order to minimize the amplitude of surface motion, the optimized strategies are keeping the lengths of the jets small, using a pulsation dampener, reducing the flow rate, and adjusting mass concentration to an optimal value. Nevertheless, a high flow rate is typically beneficial for laser-driven ion acceleration: the size of the sheet is larger, and the thickness is smaller(28). Therefore, a trade-off has to be made based on the specific requirements of applications. Lastly, under the condition of a small flow rate and a high liquid viscosity, the motion amplitudes of the sheets are slightly smaller for a collision angle of 90° over 60°.

A comparison between the motion amplitudes of the sheets and the Rayleigh length of the laser pulse is of particular interest. Currently, the peak power of kHz-repetition-rate femtosecond lasers is typically less than 1 TW. In order to achieve a high intensity for ion acceleration, the laser pulses typically are tightly focused with a small F-number (F/#) of ~1. The diffraction limited ($1/e^2$) focal spot radius $\omega_0$ is calculated as

$$\omega_0 = 1.22\lambda \cdot F,$$

where $\lambda$ is the wavelength of the laser and it takes the value of 800 nm. Therefore, the focal spot radius $\omega_0$ ~ 1 μm. Rayleigh length $z_R$ is defined as

$$z_R = \frac{\pi \omega_0^2}{\lambda},$$

hence the corresponding Rayleigh length $z_R$ ~ 3.7 μm. According to Figure 5, the motion amplitudes in the stable region for most groups are less than 3.7 μm. For better performance, for example, <1 μm, one needs to consider adopting the proper parameters studied in our work.

**4.2  Conclusion**

In summary, we produce ethylene glycol liquid sheets in the double-jet collision scheme and perform a thorough study on their motion in the surface normal direction. We find that it is crucial to avoid the development of instabilities by optimizing the sheet generation parameters to minimize the amplitude of the motion. After optimizations, we can achieve amplitudes smaller than 1.5 µm to fully meet the requirement of sheet position stability for a tight-focusing setup in laser-ion acceleration experiments. Table-top ion and neutron sources based on kHz femtosecond TW lasers and liquid sheet targets are feasible and promising.

## 5 Conflict of Interest

*The authors declare that the research was conducted in the absence of any commercial or financial relationships that could be construed as a potential conflict of interest.*

## 6 Author Contributions

WM provided overall leadership and oversight for the experiment. ZC and ZPe conceived, conducted the experiments, performed analysis, and prepared the manuscript. YS, JZ, SC, YG, JL, PW, ZM, ZPa, DK, GQ, SRX, ZL, YL, SXX, TS, XC, QW and XL all contributed to the development of the liquid-sheet-generation system. WM helped with writing, proofreading the manuscript and reformed its structure. All authors reviewed, edited, and contributed to the manuscript.

## 7 Funding

NSFC Innovation Group Project (grant number 11921006), National Grand Instrument Project (grant number 2019YFF01014402), the National Natural Science Foundation of China (grant number 12205008) and the National Science Fund for Distinguished Young Scholars (12225501).

## 8 Acknowledgments

This work was supported by the following projects: NSFC Innovation Group Project (grant number 11921006), National Grand Instrument Project (grant number 2019YFF01014402) and the National Natural Science Foundation of China (grant number 12205008). W. Ma acknowledges support from the National Science Fund for Distinguished Young Scholars (12225501). The authors thank Prof. Ke Xu, Dr. Wei Yang, Shuai Zheng, and Fei Yu for advice on liquid-sheet-generation system design. The picture of the beaker in Figure 1B is provided by "OpenClipart-Vectors" from "pixabay.com". The content of the manuscript have been uploaded to arXiv as a preprint.

## 9 Data Availability Statement

Datasets are available on request:

The raw data supporting the conclusions of this article will be made available by the authors, without undue reservation.

## 11  Figure captions

Figure 1. (**A**) Schematic of the liquid sheets generation process in the two-jet-collision regime. (**B**) Schematic of our liquid-delivery subsystem. (**C**) Photograph of the generated liquid sheet. (**D**) Photograph of the capillary positioning subsystem.

Figure 2. (**A**) Schematic and (**B**) photograph of the measurement setup. The confocal detector is used to measure the motion of the sheets. Two imaging systems record the experimental parameters. Multiple light sources are used for imaging illumination.

Figure 3. (**A**) The arrangement of the measurement points in a typical experiment. (**B**) Raw data of the motion of a liquid sheet corresponding to one measurement at a certain point. (**C**) A zoom-in of (**B**). (**D**) The motion amplitude of the sheet as a function of the distance to the collision point. The inset in d is the typical histogram of one measurement, where two vertical lines indicate the threshold $X_{95}$.

Figure 4. Motion amplitude of liquid sheets with different generation parameters: (**A**) flow rate, (**B**) mass concentration of EG solution, (**C**) lengths of the jets, (**D**) status of pulsation dampener, (**E**) collision angle ($2\theta$).

Figure 5. A summary of sheet generation parameters' influences on the motion amplitude of the liquid sheet.